# Initial Performance of BICEP3: A Degree Angular Scale 95 GHz Band Polarimeter

W. L. K. Wu[1,2] · P. A. R. Ade[3] · Z. Ahmed[1,2] · K. D. Alexander[4] ·
M. Amiri[5] · D. Barkats[4] · S. J. Benton[6] · C. A. Bischoff[4] · J. J. Bock[7,8] ·
R. Bowens-Rubin[4] · I. Buder[4] · E. Bullock[9] · V. Buza[4] · J. A. Connors[4] ·
J. P. Filippini[10] · S. Fliescher[9] · J. A. Grayson[1,2] · M. Halpern[5] ·
S. A. Harrison[4] · G. C. Hilton[11] · V. V. Hristov[7] · H. Hui[7] · K. D. Irwin[1,2] ·
J. Kang[1,2] · K. S. Karkare[4] · E. Karpel[1,2] · S. Kefeli[7] · S. A. Kernasovskiy[1,2] ·
J. M. Kovac[4] · C. L. Kuo[1,2] · K. G. Megerian[8] · C. B. Netterfield[6] ·
H. T. Nguyen[8] · R. O'Brient[7,8] · R. W. Ogburn IV[1,2] · C. Pryke[9] ·
C. D. Reintsema[11] · S. Richter[4] · C. Sorensen[4] · Z. K. Staniszewski[7,8] ·
B. Steinbach[7] · R. V. Sudiwala[3] · G. P. Teply[7] · K. L. Thompson[1,2] ·
J. E. Tolan[1,2] · C. E. Tucker[3] · A. D. Turner[8] · A. G. Vieregg[12] ·
A. C. Weber[8] · D. V. Wiebe[6] · J. Willmert[9] · K. W. Yoon[1,2]

**Abstract** BICEP3 is a 550-mm aperture telescope with cold, on-axis, refractive optics designed to observe at the 95-GHz band from the South Pole. It is the newest member of the BICEP/*Keck* family of inflationary probes specifically designed to measure the polarization of the cosmic microwave background (CMB) at degree angular scales. BICEP3 is designed to house 1280 dual-polarization pixels, which, when fully populated, totals to ∼9× the number of pixels in a single *Keck* 95-GHz receiver, thus further advancing the BICEP/*Keck* program's 95 GHz mapping speed. BICEP3 was deployed

✉ W. L. K. Wu
wlwu@stanford.edu

1 Department of Physics, Stanford University, Stanford, CA 94305, USA

2 Kavli Institute for Particle Astrophysics and Cosmology, SLAC National Accelerator Laboratory, Menlo Park, CA 94025, USA

3 School of Physics and Astronomy, Cardiff University, Cardiff CF24 3AA, UK

4 Harvard-Smithsonian Center for Astrophysics, Cambridge, MA 02138, USA

5 Department of Physics, University of Toronto, Toronto, ON, Canada

6 Department of Physics and Astronomy, University of British Columbia, Vancouver, BC, Canada

7 Division of Physics, Mathematics and Astronomy, California Institute of Technology, Pasadena, CA 91125, USA

8 Jet Propulsion Laboratory, Pasadena, CA 91109, USA

during the austral summer of 2014–2015 with nine detector tiles, to be increased to its full capacity of 20 in the second season. After instrument characterization, measurements were taken, and CMB observation commenced in April 2015. Together with multi-frequency observation data from Planck, BICEP2, and the *Keck Array*, BICEP3 is projected to set upper limits on the tensor-to-scalar ratio to $r \lesssim 0.03$ at 95 % C.L.

**Keywords** Cosmic microwave background · Primordial gravitational waves · Inflation · Instrumentation: polarimetry · Telescopes

## 1 Introduction

Inflation is the leading model that explains the spatial flatness of the universe, provides a mechanism that generates the initial conditions for structure formation, and explains the uniformity in the temperature of the cosmic microwave background (CMB). It generically predicts a stochastic gravitational wave background (GWB) from stretching the tensor perturbations of the metric during the exponential expansion. This GWB is the only source that generates a curl polarization pattern—the 'B-modes'—on the CMB at recombination (see, e.g., [1], for references). The ratio of the amplitude of the tensor perturbation spectrum to the scalar perturbation spectrum, the tensor-to-scalar ratio $r$, is directly related to the energy scale of inflation. Therefore, detection of B-modes generated by primordial gravitational waves is key to understanding the inflationary epoch.

However, polarized thermal dust emission from our galaxy also generates B-mode polarization (see, e.g., [2]). BICEP2 reported a detection of B-modes at 150 GHz at degree angular scales at a level equivalent to $r = 0.2$ [1]. Joint analysis of BICEP2/*Keck Array*'s 150 GHz data with Planck's multi-frequency data detected polarized dust at high significance and only set an upper limit on $r$ to $r < 0.12$ at 95 % C.L., limited by noise uncertainties at frequencies other than 150 GHz [3]. In order to disentangle the dust B-mode amplitude from $r$, we need observations at frequencies other than 150 GHz with map depth similar to BICEP2/*Keck Array*'s 150 GHz maps at 3.0 µK·arcmin, as of 2014 [4].

To this end, the BICEP/*Keck* family of telescopes are observing at multiple frequencies: The *Keck Array* has five BICEP2 style cameras currently observing at 95, 150, and 220 GHz. BICEP3 will add mapping speed at 95 GHz by housing 1280 dual-polarization pixels (~9× more than a single *Keck* 95-GHz receiver). The map depth achieved by two *Keck* receivers observing at 95 GHz during the 2014 season is 7.62 µK·arcmin [4]. Assuming similar observing efficiency and detector yield in the fully populated BICEP3 focal plane with 1280 pixels, BICEP3 alone can reach a map depth of 3.6 µK·arcmin from its first full season of observation.

9 School of Physics and Astronomy, University of Minnesota, Minneapolis, MN 55455, USA
10 Department of Physics, University of Illinois at Urbana-Champaign, Urbana, IL 61820, USA
11 National Institute of Standards and Technology, Boulder, CO 80305, USA
12 Department of Physics, Enrico Fermi Institute, University of Chicago, Chicago, IL 60637, USA

## 2 Instrument Overview

BICEP3 adopts the design strategy of its predecessors. It has cold, on-axis, refractive optics with a 550-mm aperture, and it observes from the South Pole on the three-axis mount designed for BICEP1. More detailed descriptions of the instrument can be found in [5,6], and here we summarize the various subsystems of the telescope.

BICEP3 adopts the same detector technology as BICEP2/*Keck*—beam-defining slot-antenna arrays, their summing networks, band-defining filters, and the transition-edge sensors (TES) are photolithographed on a silicon substrate to form the detector tile [7]. The TESs are read out at 270 mK by superconducting quantum interference devices (SQUIDs) via a time-domain multiplexing scheme. In BICEP3, rather than having MUX chips and their associated readout circuits laid out on the perimeter of the focal plane area as in BICEP2 and *Keck*, we package the readout electronics behind the detector tile in detector modules. Each module contains 64 detector pairs. This modularization makes more efficient use of the focal space for detectors in BICEP3 and allows us to pack 20 detector tiles (Fig. 1).

Alumina is used as the lens material for its higher refractive index ($n = 3.11$, about twice that of HDPE), which allows the lenses to be thinner (2.1 and 2.7 cm at the center, respectively) and therefore reduces image aberrations. Alumina also has $\mathcal{O}(100)$ better thermal conductivity than HDPE. This reduces the loading on the detectors by having a smaller temperature gradient across the lens (Fig. 1).

Both absorptive and reflective infrared filters are employed in BICEP3 to reject out-of-band loading. For absorptive filters, we have an alumina filter at the top of the 50 K stage and two nylon filters in the 4 K optics tube. For reflective filters, we use large area metal-mesh capacitive filters with frequency cutoffs around a few THz custom designed for BICEP3 [8]. These are placed behind the window to reject at least 80 % of the incoming 165 W of infrared radiation, so as to not overload the pulse tube cooler which has ∼40 W of cooling power at the 50 K stage.

The receiver cryostat consists of three nested aluminum cylinders at room temperature, 50 and 4 K, with the 50 and 4 K stage cooled by a pulse tube cooler Cryomech PT415. The focal plane is attached to the ultra-cold stage of the sub-K structure, which is cooled by a 3-stage (He4/He3/He3) helium sorption fridge from Chase Research Cryogenics.

## 3 Operations and Initial Performance

We deployed BICEP3 in the Dark Sector Lab (DSL) at the South Pole, Antarctica, during the austral summer of 2014–2015. Nine detector modules, with average 67 % yield, were installed in its camera for the 2015 season of observation. The focal plane will be fully populated with new, higher-yield detector tiles in the 2016 season. We have made instrument and detector characterization and optimization measurements including near-field and far-field beam responses, detector spectral response, magnetic response, detector optical efficiency, detector thermal loading measurements, cryostat thermal loading characterizations, and detector readout optimization. We started scanning the same 1000 sq. degree CMB field centered at RA = 0 h and dec = $-57.5°$ with the same scan pattern as BICEP2 and *Keck* [9]. Here we present preliminary results of the instrument characterization and CMB maps with preliminary sensitivity estimates.

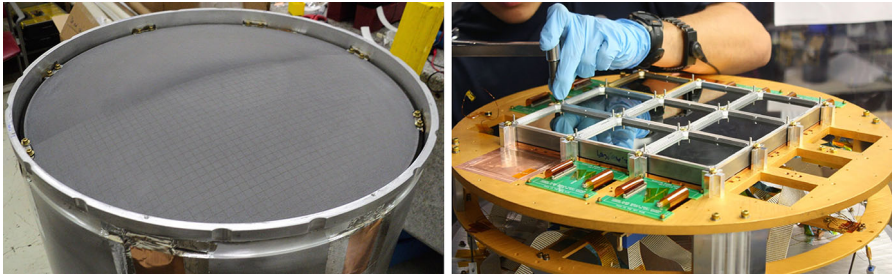

**Fig. 1** *Left* alumina objective lens with epoxy-based AR coating diced at 1-cm pitch for strain relief installed in 4 K optics tube. *Right* BICEP3 focal plane with 9 detector modules installed, to be increased to 20 detector modules for the 2016 observing season (Color figure online)

### 3.1 Beams

The far-field response of each detector is measured in order to characterize the level at which beam systematics enter the polarization measurements. To measure the far-field response of the detectors, we use a chopped thermal source mounted on a mast on the roof of MAPO (∼200 m from DSL, close to BICEP3's far-field). We mount a flat mirror above the BICEP3 window so that the source's radiation is directed toward the detectors. The source chops between the sky (∼10 K) and ambient temperature (∼250 K) at a rate of 18 Hz. The detector time streams are then demodulated to form the beam maps. An array-averaged beam map is shown in Fig. 2. We extract the *x* and *y* positions, beam width, and ellipticities of the beams from the beam maps. Most relevant for systematics in the polarization signal are the differential beam parameters, which could be extracted from differencing the elliptical Gaussian fits to the beam maps of A and B detectors in a pair.

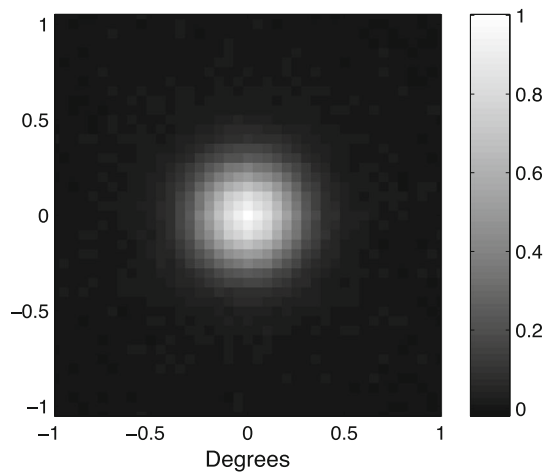

**Fig. 2** Array-averaged measured far-field beam pattern of BICEP3 detectors, linear scale

**Table 1** Differential beam parameters: median, detector-to-detector scatter, and median measurement uncertainty

| Parameter | Median | Scatter | Measurement uncertainty |
|---|---|---|---|
| Differential pointing, $\delta x$ (arcmin) | 0.11 | 0.12 | 0.03 |
| Differential pointing, $\delta y$ (arcmin) | −0.62 | 0.16 | 0.05 |
| Differential beam width (°) | 0.0057 | 0.001 | 0.000 |
| Differential ellipticity '+' | −0.018 | 0.013 | 0.004 |
| Differential ellipticity '×' | −0.006 | 0.008 | 0.003 |

Preliminary analysis of the far-field beam maps shows that the median beam width, $\sigma$ of the Gaussian fit, of all the BICEP3 detectors is $0.171° \pm 0.005°$. Table 1 presents the differential beam parameters from this analysis. Compared to BICEP2 [10], the differential pointing parameters $\delta x$ and $\delta y$ are smaller; the differential beam width is larger, but is at a low enough level that can be cleaned by deprojection [11]; the measurement uncertainties of differential ellipticities are smaller, which is important because differential ellipticities are subtracted as opposed to deprojected.

### 3.2 Detector Spectral Response

We design BICEP3 to observe at a band centered around 93 GHz with a fractional bandwidth of 25 %. Therefore, the detector slot antennas, band-defining filters, optics, and their AR coatings are tuned corresponding to this frequency band. This design effectively avoids the oxygen lines at 118 GHz on the high end of the band pass and a family of oxygen lines at 56.3, 58.45, and 62.5 GHz on the low end.

FTS data are collected by a custom-built Martin–Puplett interferometer that is mounted above the window of BICEP3 [12]. The resulting interferograms are filtered and then Fourier transformed to obtain the frequency response of the detectors.

Preliminary analysis of the FTS data shows that the array band centers are at 90.4 GHz with bandwidth of 24.7 GHz from seven detector tiles that have enough FTS pointings out of the nine deployed. The bandwidth is within design expectations, and the band center is a few GHz lower from the detector design goal of 93 GHz. However, this combination of band center and bandwidth is safe from any of the atmospheric bright lines. Figure 3 shows the co-added frequency response spectrum from all available detectors.

### 3.3 CMB Map

The CMB observation data that BICEP3 collected from mid-April to mid-July is analyzed to make preliminary maps and to estimate the NET of the instrument. Approximate pointing is used, relative gain calibration is derived from elevation nod response,

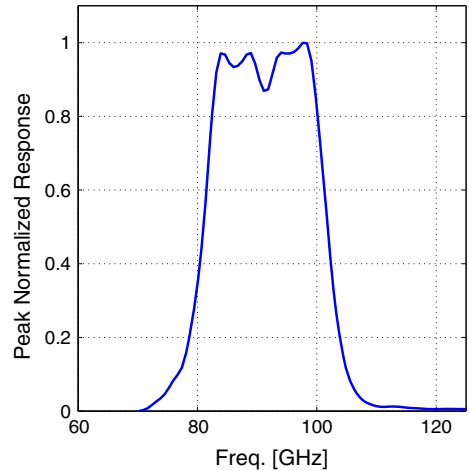

Fig. 3 The array-averaged frequency response spectrum. The band center is at 90.4 GHz with bandwidth of 24.5 GHz (Color figure online)

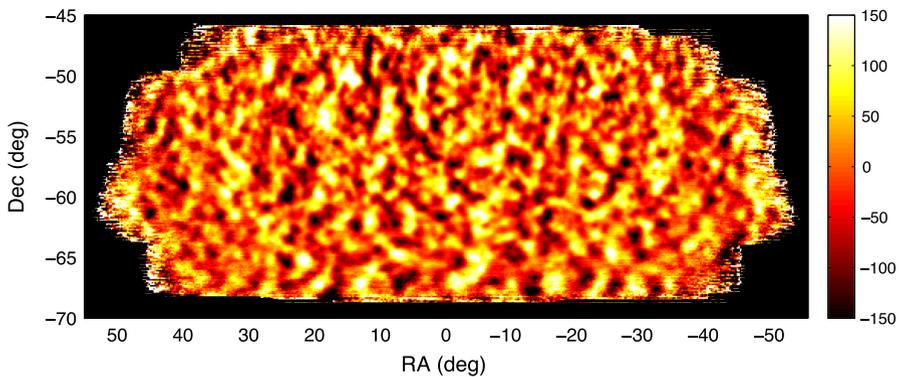

Fig. 4 Preliminary map from BICEP3 observations of the primary CMB field. This temperature map is made with 900 h of integration on source, from April 20 through July 13. The color scale is in µK (Color figure online)

the cut structures are inherited from BICEP2/*Keck* analysis, and cut thresholds are preliminary. Figure 4 shows the temperature map. The fluctuations are in good agreement with the CMB field as imaged by BICEP2/*Keck*.

From this map, we estimate the per-detector NET to be about 395 µK·$\sqrt{s}$, 60 % higher than *Keck*'s 95 GHz detectors. We attribute a large proportion of the higher NET to excess optical loading in the BICEP3 optics train. The exact breakdown of the various sources of extra loading is being analyzed at the time of writing.

## 4 Summary and Looking Forward

BICEP3, a new telescope of the BICEP/*Keck* program designed to observe at the 95-GHz band, was deployed at the South Pole during the austral summer of 2014–2015. Instrument characterization measurements were taken, and CMB observations are underway. We have given the available data a preliminary analysis and found that the beam shapes, the detector spectral response, and temperature map features are within expectations. More instrument characterization measurements will be performed in the coming austral summers, and hardware upgrades are planned. In particular, we will field 20 detector modules in the 2015–2016 summer season to fully populate the focal plane. With additional improvements and optimizations on loading and readout performance, we project the array NET to reach 6.9 $\mu K \cdot \sqrt{s}$.

**Acknowledgments** This work is supported by the National Science Foundation (Grant Nos. 1313158, 1313010, 1313062, 1313287, 1056465, 0960243), the SLAC Laboratory Directed Research and Development Fund, the Canada Foundation for Innovation, Science and Technology Facilities Council Consolidated Grant (ST/K000926/1), and the British Columbia Development Fund. The development of detector technology was supported by the JPL Research and Technology Development Fund and Grants 06-ARPA206-0040, 10-SAT10-0017, and 12-SAT12-0031 from the NASA APRA and SAT programs. The development and testing of detector modules were supported by the Gordon and Betty Moore Foundation.

# References

1. BICEP2 Collaboration, Phys. Rev. Lett. **112**, 241101 (2014), arXiv:1403.3985
2. J. Dunkley, A. Amblard, C. Baccigalupi, M. Betoule, D. Chuss, A. Cooray, J. Delabrouille, C. Dickinson, G. Dobler, J. Dotson, et al., in *American Institute of Physics conference series*, vol. 1141 (2009), pp. 222–264, arXiv:0811.3915
3. BICEP2/Keck and Planck Collaborations, Phys. Rev. Lett. **114**, 101301 (2015), arXiv:1502.00612
4. Keck Array and BICEP2 Collaborations, P.A.R. Ade, Z. Ahmed, R.W. Aikin, K. D. Alexander, D. Barkats, S.J. Benton, C.A. Bischoff, J.J. Bock, R. Bowens-Rubin et al., ArXiv e-prints (2015), arXiv:1510.09217
5. Z. Ahmed, M. Amiri, S.J. Benton, J.J. Bock, R. Bowens-Rubin, I. Buder, E. Bullock, J. Connors, J.P. Filippini, J.A. Grayson et al., in *Society of Photo-Optical Instrumentation Engineers (SPIE) Conference Series*, vol. 9153 (2014), p. 1, arXiv:1407.5928
6. W.L.K. Wu, Ph.D. thesis, Stanford University (2015)
7. P.A.R. Ade, R.W. Aikin, M. Amiri, D. Barkats, S.J. Benton, C.A. Bischoff, J.J. Bock, J.A. Bonetti, J.A. Brevik, I. Buder et al., ArXiv e-prints **1502**, 00619 (2015)
8. Z. Ahmed, J.A. Grayson, K.L. Thompson, C.-L. Kuo, G. Brooks, T. Pothoven, J. Low Temp. Phys. **176**, 835 (2014), arXiv:1407.5892
9. BICEP2 Collaboration, ApJ **792**, 62 (2014), arXiv:1403.4302
10. BICEP2 and Keck Array Collaborations, ApJ **806**, 206 (2015), arXiv:1502.00596
11. BICEP2 Collaboration, ArXiv e-prints **1502**, 00608 (2015)
12. K.S. Karkare, P.A.R. Ade, Z. Ahmed, R.W. Aikin, K.D. Alexander, M. Amiri, D. Barkats, S.J. Benton, C.A. Bischoff, J.J. Bock, et al., in *Society of photo-optical instrumentation engineers (SPIE) conference series*, vol. 9153 (2014), p. 3